\documentclass[10pt,letterpaper]{article}
\usepackage{opex3}
\usepackage{graphicx}  
\begin{document}

\title{Ramsey interferometry with an atom laser}

\author{D D\"oring$^{1,2}$, J E Debs$^{1,2}$, N P Robins$^{1,2}$, C Figl$^{1,2}$, P A Altin$^{1,2}$ and J D Close$^{1,2}$}

\address{$^1$ Australian Research Council Centre of Excellence for Quantum-Atom Optics, The Australian National University, ACT 0200, Australia}
\address{$^2$ Department of Quantum Science, The Australian National University, ACT 0200, Australia}
\email{daniel.doering@anu.edu.au}

\begin{abstract}
We present results on a free-space atom interferometer operating on the first order magnetically insensitive $\left|F=1,m_F=0\right\rangle\rightarrow\left|F=2,m_F=0\right\rangle$ ground state transition of Bose-condensed $^{87}$Rb atoms. A pulsed atom laser is output-coupled from a Bose-Einstein condensate and propagates through a sequence of two internal state beam splitters, realized via coherent Raman transitions between the two interfering states. We observe Ramsey fringes with a visibility close to $100\,\%$ and determine the current and the potentially achievable interferometric phase sensitivity.  This system is well suited to testing recent proposals for generating and detecting squeezed atomic states. 
\end{abstract}

\section{Introduction}
Atom interferometry \cite{Cronin2007} has proven to be an increasingly valuable technique for precision measurements over the last years \cite{J2007, G2008, S2002, Malo2008, Jan2008}. Compared to photons, atoms offer the advantage of having an intrinsically more complex structure and therefore allowing a larger range of possible measurements to be undertaken. There have been a number of fundamentally important experiments making use of the atomic mass to measure the Newtonian gravitational constant $G$ \cite{J2007,G2008} and the fine structure constant $\alpha$ \cite{S2002,Malo2008}. Recently, atomic interferometers have been proposed to be used for tests of general relativistic effects \cite{Kasevich1} and the detection of gravitational waves \cite{Kasevich2}. All of these experiments and proposals are based on thermal atoms. In many cases, such as atomic Sagnac interferometry \cite{T2007}, improved performance can be achieved by using laser-cooled atoms from a magneto-optical trap, substantially narrowing the atomic velocity distribution. Bose-Einstein condensates (BECs) comprise a macroscopic number of atoms in a single momentum state and have an even narrower momentum width, limited by the Heisenberg uncertainty principle. This narrow velocity spread makes BECs an excellent candidate for highly velocity selective light-based beam splitters in atom interferometers. Furthermore, Bose-condensed atoms in the form of a freely propagating pulse, known as an atom laser, are a promising way to produce reduced-uncertainty states (squeezed states) \cite{D1994,ueda,S2005,Mattias2007}, potentially leading to a further increase in interferometric sensitivity (analog to the increased phase sensitivity in optical interferometers by using squeezed photon states \cite{caves1981}).  

Atom interferometry with confined high density BECs is strongly affected by phase diffusion due to mean field interactions in the condensate, significantly limiting the coherence time \cite{Cronin2007,Castin1997}. Experiments to overcome this barrier have been conducted by, e.g., reducing the atomic interactions via a magnetic Feshbach resonance \cite{fattori2008} or utilizing number squeezed states \cite{Kasevich3,Jo2007}. Alternatively, mean field interactions can be significantly lowered by using a more dilute cloud of atoms. Releasing the BEC from the confining potential and thereby decreasing its density greatly reduces the issue of phase diffusion and makes Bose-condensed atoms a promising alternative for free-space atom interferometry, as demonstrated by Torii \textit{et al.} \cite{Yoshio2000} in the context of Bragg diffraction interferometry with atomic momentum states.

Internal state atom interferometers commonly rely on Ramsey-type interferometry, using the method of separated oscillatory fields \cite{Norman1950}. The most fundamental version of this technique includes two temporally or spatially separated beam splitters that couple a two-level system to an external driving field. Control of the time between the two beam splitters or the phase or the frequency of the driving field allows the observation of Ramsey fringes -- an oscillation in population transfer between two internal states. The position of the fringe pattern contains information about the relative phase accumulated by the two atomic states, enabling precision measurements with atom inteferometers. For uncorrelated atoms, the precision of such a device is limited by the projection noise in the detection process \cite{D1994}, commonly known as the standard quantum limit. To achieve detection sensitivities beyond this limit, squeezed states \cite{ueda} have to be used as a source for the interferometer. Spin squeezing has been demonstrated with thermal Rubidium \cite{Monika2008} and Cesium \cite{J2008} atoms, yielding reduced fluctuations in the population difference between the two internal states used in modern atomic fountain clocks. Recently, Est$\grave{\mbox{e}}$ve \textit{et al.} \cite{JE2008} achieved squeezing of Bose-condensed atoms in an optical lattice. The decreased uncertainty in the relative number of atoms at neighboring lattice sites makes their setup a good candidate for improved precision interferometric measurements. Different mechanisms for squeezing of a freely propagating atom laser have been proposed by Haine \textit{et al.} \cite{S2005} and Johnsson \textit{et al.} \cite{Mattias2007}. The setup presented in this Letter is well suited for the implementation of either of these schemes and the most direct way of detecting the effect of squeezing on the performance of an atom laser interferometer.

\section{The atom laser interferometer}

We here present results on a free-space $^{87}$Rb atom laser interferometer operating on the $\left|F=1,m_F=0\right\rangle\rightarrow\left|F=2,m_F=0\right\rangle$ ground state transition, which is to first order insensitive to magnetic fields. An atom laser consisting of $N=7.7\times10^4$ atoms is output-coupled from a Bose-Einstein condensate and travels under gravity through a sequence of two Ramsey-type (internal state) beam splitters.  To the best of our knowledge, this is the first such demonstration using Bose-condensed, instead of thermal atoms, in a free-space internal-state interferometer.  

\subsection{Experimental scheme}

The setup for producing Bose-Einstein condensates is described in detail in reference \cite{D2008}. In brief, we collect and cool $10^{10}$ $^{87}$Rb atoms in a three-dimensional magneto-optical trap (MOT), loaded from a two-dimensional MOT. After polarization gradient cooling and optical pumping into the desired internal ground state $\left|F=1, m_F=-1\right\rangle$, the atoms are mechanically transported and transferred into a harmonic quadrupole-Ioffe-configuration (QUIC) trap. We use radio-frequency induced evaporative cooling and achieve a Bose-Einstein condensate of up to $10^6$ atoms in the $\left|F=1,m_F=-1\right\rangle$ state. By means of radio frequency (rf) induced spin-flips \cite{M1997}, atoms are coherently transferred from the condensate into the untrapped $\left|F=1,m_F=0\right\rangle$ state, in which they fall away from the condensate under gravity.

\begin{figure}[b]
\begin{center}
\includegraphics[scale=0.48]{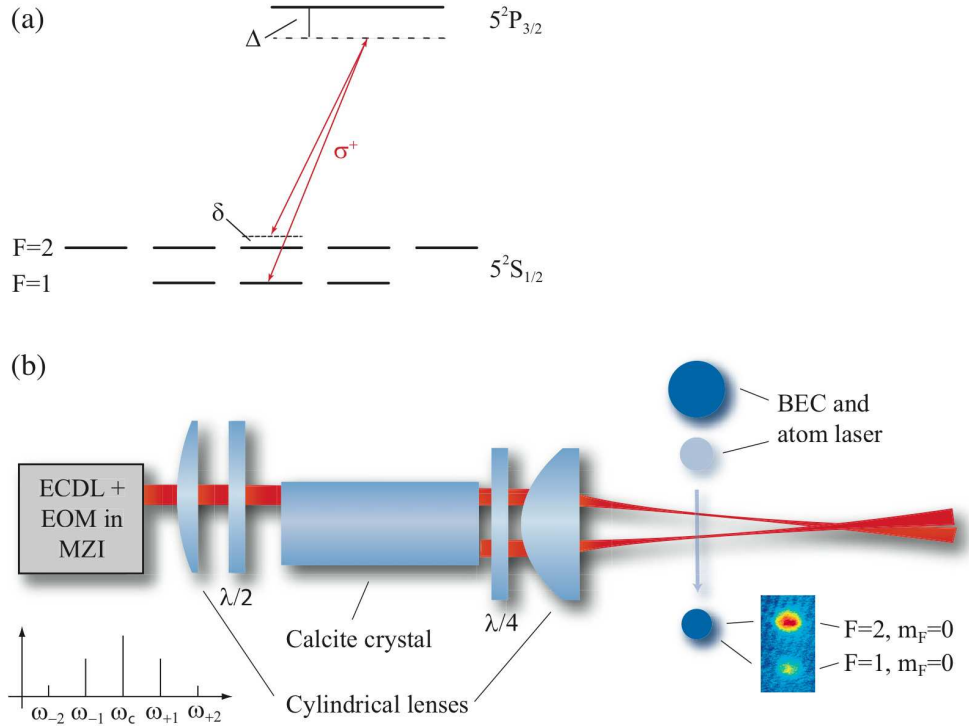}

\caption{\label{fig:setup} (a) Level scheme and (b) experimental setup of the Raman beam splitters for the atom interferometer. The atom laser pulse propagates under gravity through the two beam splitters and is subsequently imaged via a state selective absorption imaging scheme. Splitting the optical beam path with a calcite crystal close to the atom laser interferometer allows for improved relative stability of the two beam splitter positions. The light source is an external cavity diode laser (ECDL), in combination with an electro-optic modulator (EOM) placed in one arm of a Mach-Zehnder interferometer (MZI).}
\end{center}
\end{figure}

Starting from the output-coupled $\left|F=1,m_F=0\right\rangle$ atom laser pulse, we construct an internal state interferometer for the two ground states $\left|F=1, m_F=0\right\rangle$ and $\left|F=2, m_F=0\right\rangle$ of $^{87}$Rb. They are coupled by copropagating circularly polarized focussed Raman beams, allowing spatially selective addressing of the atom laser. In order to achieve the required high relative phase stability of the different frequency components in the Raman beams, we use a single external cavity diode laser (ECDL), detuned by $\Delta=100\,$GHz from the Rb $D_2$-line, as the laser source for the interferometer. Via an electro-optic phase modulator (EOM), we produce sidebands with a spacing of $3.4\,$GHz, corresponding to half the ground state hyperfine splitting of $^{87}$Rb, on the light coming from the ECDL. The stability of the microwave source is improved by locking it to a GPS-based external reference oscillator. As the Rabi frequency for Raman transitions driven by purely phase-modulated light vanishes due to destructive interference \cite{JD2008,P2003}, the EOM is placed in one arm of a Mach-Zehnder interferometer (MZI). This allows us to partly convert phase into amplitude modulation and drive the desired transition in the atom laser beam splitters. To reduce phase fluctuations between the two frequency components driving the Raman transitions, the two arms of the Mach-Zehnder interferometer are locked using a piezo-controlled mirror and a balanced homodyne detection scheme. The modulation setup is similar to the one described in \cite{JD2008}, where a true two-state atom laser output-coupling scheme is achieved via selective Raman transitions between only two states of the ground state hyperfine manifold \cite{JD2008,Immanuel2001}. 

The geometric design for the realization of two spatially separated Ramsey beam splitters is shown in figure~\ref{fig:setup}. Two vertically displaced sheets of light are produced using a combination of a calcite crystal and cylindrical lenses. The light from the modulation setup is pre-focussed by the first cylindrical lens (focal length $f=500\,$mm) and subsequently sent through a birefringent calcite crystal to produce two beams with a spatial separation of $2\,$mm. The second cylindrical lens ($f=100\,$mm) focusses the light onto the propagation axis of the atom laser and generates two sheets of light with a width ($2w_0$) of $40\,\mu$m. A pulsed atom laser is output-coupled from the condensate and travels through the Raman light sheets. The velocity of the atom laser pulse depends on its fall time, which means that for an identical splitting ratio, the intensity requirements for the upper and the lower beam splitter are different. The relative intensity is controlled by a $\lambda/2$-waveplate placed before the (polarization-dependent) calcite crystal. The propagation time $T_p=6.04(3)\,$ms between the beam splitters is determined from the Ramsey fringe pattern (see below). In a separate calibration experiment, we measure the time $T_0$ it takes to reach the first beam splitter, $T_0=5.0(2)\,$ms. Converting these measurements into spatial coordinates yields a separation of $474(10)\,\mu$m between the two beam splitters and a distance of $123(10)\,\mu$m between the center of the Bose-Einstein condensate and the upper beam splitter. Relative position fluctuations of the two light sheets are strongly supressed in this system, as the beams travel along the same path up to the calcite crystal, and vibrations of optical elements in the split beam path are of common mode to both of the beams. A stable absolute position is ensured by a rigid mechanical assembly.

After traversing the two beam splitters, the atom laser pulse is imaged via a state selective absorption imaging sequence. We first image atoms in the $F=2$ hyperfine state with a resonant light pulse. The momentum imparted by the imaging photons is large enough for the atoms to leave the imaging region within a time well below $1\,$ms. After a delay time of $1\,$ms, corresponding to a travel distance of the atom pulse of about $150\,\mu$m, we repump all atoms from the $F=1$ into the $F=2$ manifold with a light pulse resonant to the $F=1$ atoms and subsequently take a second absorption image. This procedure yields an independent measurement of the population in both internal states, allowing us to neglect the influence of run-to-run number fluctuation in our Bose-Einstein condensates.

\subsection{Ramsey fringe measurement}   

\begin{figure}[h]
\begin{center}
\includegraphics[scale=0.85]{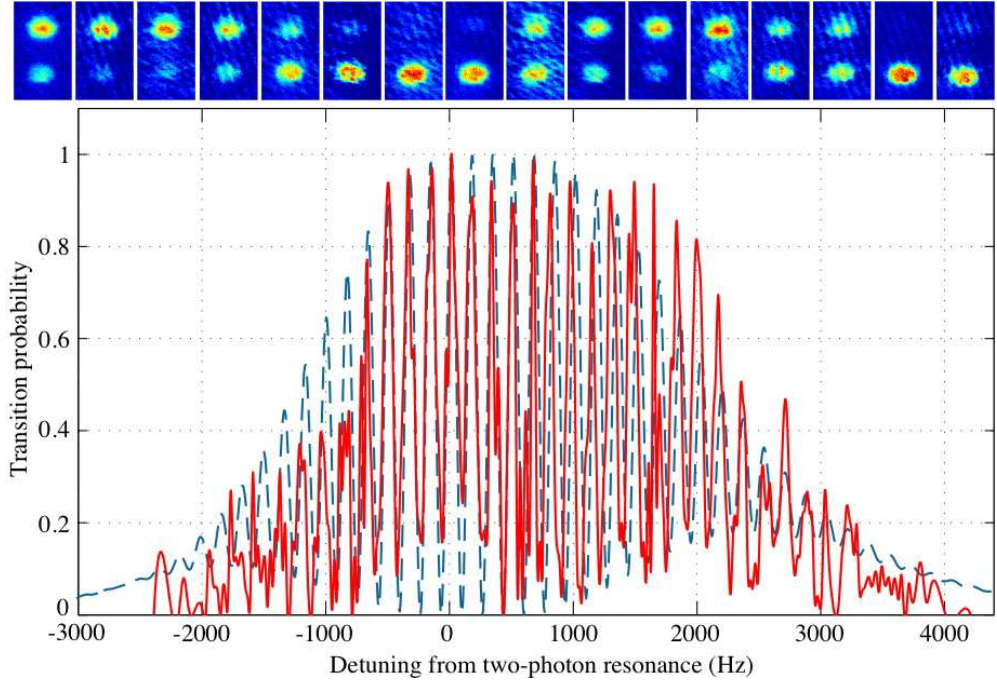}
\end{center}
\caption{\label{fig:fringes} Ramsey fringes measured over a range of $6.5\,$kHz. The red solid curve shows an experimental data set, and the blue dashed curve depicts simulated Ramsey fringes for comparable experimental conditions. The upper (lower) clouds of atoms in the absorption pictures above the graph show the population in the $\left|F=2,m_F=0\right\rangle$ ($\left|F=1,m_F=0\right\rangle$) state, for different detunings from two-photon resonance. Two adjacent absorption pictures correspond to a frequency difference of $20\,$Hz.}
\end{figure}

We record Ramsey fringes by varying the detuning $\delta$ from two-photon resonance with the EOM. The solid red line in figure~\ref{fig:fringes} depicts a typical experimental data set. The Ramsey fringe period is determined via a fit to the central part of the fringe pattern and gives a value of $f_0=165.7(8)\,$Hz. The propagation time between the beam splitters is the inverse of that value, $T_p=6.04(3)\,$ms. The power-broadened width of the Ramsey fringe envelope (FWHM) is about $3\,$kHz, determined by the two-photon Rabi frequency. We adjust the light intensity such that for a single beam splitter $50\,\%$ of the atomic population is transferred, and thus the envelope width corresponds to the inverse of the traversal time through the interaction zones. The visibility of the Ramsey fringes is close to $100\,\%$, showing the good spatial mode matching and narrow velocity distribution of the interfering atom pulse. We apply a semi-classical model for the light-atom interaction and simulate the Ramsey interferometer based on Raman transitions between the two internal states driven by a light field containing two frequencies.  The result of the simulation with conditions comparable to the ones present in the experiment is shown by the blue dashed line in figure~\ref{fig:fringes}. This simplified theoretical simulation shows reasonable agreement with the shape of the Ramsey fringe pattern. Due to differential light shifts, the Ramsey fringes are not centered around zero two-photon detuning $\delta$.

\begin{figure}[h]
\begin{center}
\includegraphics[scale=0.75]{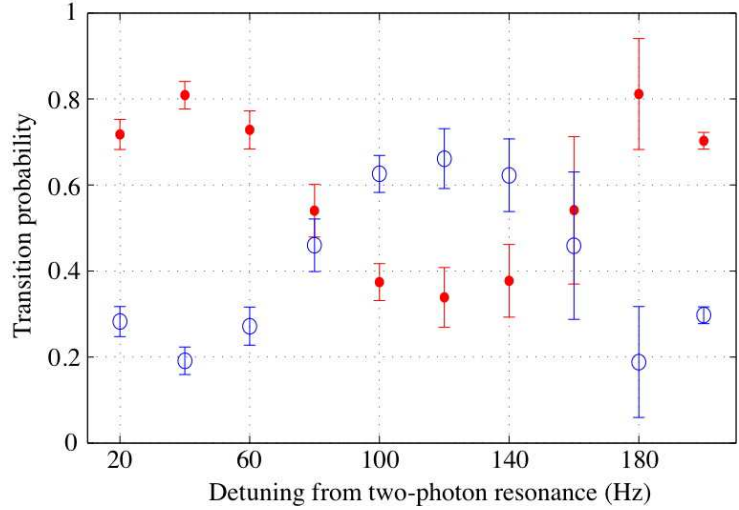}
\end{center}
\caption{\label{fig:signal-to-noise} Noise measurement for a typical sequence of Ramsey fringes. Each data point is averaged over five experimental runs. The red dots (blue circles) show the population in the $\left|F=2,m_F=0\right\rangle$ ($\left|F=1,m_F=0\right\rangle$) state.}
\end{figure}

\subsection{Interferometric sensitivity}

The phase sensitivity of the interferometer is determined by the signal-to-noise ratio on the Ramsey fringes. We determine the noise via a separate measurement of a Ramsey fringe (figure~\ref{fig:signal-to-noise}). For each data point in the graph, we average over five experimental runs, corresponding to a measurement time of six minutes per point. Given the uncertainties (SEM) in figure~\ref{fig:signal-to-noise} and the measured fringe visibility, we calculate the minimum measurable phase shift to $0.24(10)\,$rad. The accumulated phase in an interferometer is usually proportional to the measured quantity, such as a rotation or a differential energy shift due to a magnetic field. For demonstrative reasons, we calculate the frequency sensitivity in the presented apparatus and obtain a value of $6.3(2.6)\,$Hz, corresponding to a relative frequency sensitivity of $9.2(3.8)\times 10^{-10}$. This sensitivity does not compete with highly accurate frequency measurements in atomic clocks (see, e. g., \cite{wynands}); however, the purpose of this paper is to demonstrate an atom laser based Ramsey interferometer rather than the improvement of current frequency sensitivities. The overall number of detected atoms per shot is $N=7.7\times10^4$. Mechanical fluctuations in the imaging beam path lead to a time varying interference pattern on the images, resulting in a standard deviation of the detected atom number of $\sigma=4\times10^3$. However, due to our normalization procedure, this uncertainty does not strongly influence the observed Ramsey signal. The current sensitivity is limited by different noise sources. The Mach-Zehnder interferometer used in the light modulation setup introduces a phase difference between its two arms, and fluctuations in that parameter inevitably lead to a change in Rabi frequency and light shift at the positions of the two beam splitters. Despite our balanced homodyne locking scheme, this mechanism is a dominant source of uncertainty and strongly increases the noise on the Ramsey signal. Furthermore, mechanical vibrations of the optical components in the beam path change the position of the beam splitters compared to the position of the condensate in the magnetic trap. We reduce the influence of mechanical vibrations by using a solid and compact optical setup. In the regime that we are operating the interferometer, phase noise of the microwave source driving the electro-optic modulator can most likely be neglected. By eliminating all the above mentioned noise sources, optimizing the detected atom number and extending the propagation time $T_p$ from $6\,$ms to $65\,$ms (the geometrical limit of our current apparatus), we can potentially achieve a phase sensitivity of $9.8\times10^{-4}\,$rad (corresponding to a frequency sensitivity of $2.4\,$mHz) when operating at the standard quantum limit. 

\section{Conclusion}

In the work presented above, we demonstrate free-space Ramsey interferometry on the first order magnetically insensitive transition $\left|F=1,m_F=0\right\rangle\rightarrow\left|F=2,m_F=0\right\rangle$ of Bose-condensed $^{87}$Rb atoms. We plan to extend the current setup and build an atomic Mach-Zehnder interferometer with spatially separated arms, allowing us to introduce sensitivity to spatial effects, such as rotations, into our experiment. Due to its narrow velocity distribution, the atom laser is well suited for large momentum transfer beam splitters, which are a promising way towards increased interferometric sensitivity. However, the average flux of the atom laser is lower than that achieved in thermal atom interferometer sensors, yielding a potentially lower signal-to-noise ratio. To a large extent, this could be improved by extending the recently realized pumping experiment \cite{N2008,DD2008} towards a truly continuous atom laser. The alternative route towards a decreased variance in (relative) atom number is the introduction of squeezing onto our atom laser beam. When the major noise contribution is given by the quantum projection noise of the population in the two interfering states, squeezing of the pseudo-spin representing the effective two-level system can significantly decrease the variance. There have been different proposals of how to achieve squeezing of an atom laser beam \cite{S2005,Mattias2007}, both of which can potentially be realized in our setup.

\section{Acknowledgments}
This work was financially supported by the Australian Research Council Centre of Excellence program.


\begin{thebibliography}{99}

\bibitem{Cronin2007} A. D. Cronin , J. Schmiedmayer  and D. E. Pritchard, ``Optics and interferometry with atoms and molecules'', Rev. Mod. Phys. {\bf 81} 1051 (2009). 
\bibitem{J2007} J. B. Fixler , G. T. Foster, J. M. McGuirk and M. A. Kasevich, ``Atom Interferometer Measurement of the Newtonian Constant of Gravity'', Science {\bf315} 74  (2007).
\bibitem{G2008} G. Lamporesi, A. Bertoldi, L. Cacciapuoti, M. Prevedelli and G. M. Tino,  ``Determination of the Newtonian Gravitational Constant Using Atom Interferometry'', Phys. Rev. Lett. {\bf100} 050801 (2008).
\bibitem{S2002} S. Gupta, K. Dieckmann, Z. Hadzibabic and D. E. Pritchard, ``Contrast Interferometry using Bose-Einstein Condensates to Measure h/m and ${\rm \alpha}$'', Phys. Rev. Lett. {\bf89} 140401 (2002).
\bibitem{Malo2008} M. Cadoret, E. de Mirandes, P Clad\'e, S. Guellati-Kh\'elifa, C Schwob, F. Nez, L. Julien and F. Biraben, ``Combination of Bloch Oscillations with a Ramsey-Bord\'e Interferometer: New Determination of the Fine Structure Constant'', Phys. Rev. Lett. {\bf 101} 230801 (2008).
\bibitem{Jan2008} J. Friebe, A. Pape, M. Riedmann, K. Moldenhauer, T. Mehlst\"aubler, N. Rehbein, C. Lisdat, E. M. Rasel, W. Ertmer, H. Schnatz, B. Lipphardt and G. Grosche, ``Absolute frequency measurement of the magnesium intercombination transition'', Phys. Rev. A {\bf 78} 033830 (2008).
\bibitem{Kasevich1} S. Dimopoulos, P. W. Graham, J. M. Hogan and M. A. Kasevich, ``General relativistic effects in atom interferometry'', Phys. Rev. D {\bf 78} 042003 (2008). 
\bibitem{Kasevich2} S. Dimopoulos, P. W. Graham, J. M. Hogan, M. A. Kasevich and S. Rajendran, ``Atomic gravitational wave interferometric sensor'', Phys. Rev. D {\bf 78} 122002 (2008).
\bibitem{T2007} T. M\"uller, T. Wendrich, M. Gilowski, C. Jentsch, E. M. Rasel and W. Ertmer, ``Versatile compact atomic source for high-resolution dual atom interferometry'', Phys. Rev. A {\bf 76} 063611 (2007).
\bibitem{D1994} D. J. Wineland, J. J. Bollinger, W. M. Itano and D. J. Heinzen, ``Squeezed atomic states and projection noise in spectroscopy'', Phys. Rev. A {\bf 50} 67 (1994).
\bibitem{ueda} M .Kitagawa and M. Ueda, ``Squeezed spin states'', Phys. Rev. A {\bf 47} 5138 (1993).
\bibitem{S2005} S. A. Haine and J. J. Hope, ``Outcoupling from a Bose-Einstein condensate with squeezed light to produce entangled-atom laser beams'', Phys. Rev. A {\bf 72} 033601 (2005).
\bibitem{Mattias2007} M. T. Johnsson and S. A. Haine, ``Generating Quadrature Squeezing in an Atom Laser through Self-Interaction'', Phys. Rev. Lett. {\bf 99} 010401 (2007).
\bibitem{caves1981} C. M. Caves, ``Quantum-mechanical noise in an interferometer'', Phys. Rev. D {\bf 23} 1693 (1981).
\bibitem{Castin1997} Y. Castin and J. Dalibard, ``Relative phase of two Bose-Einstein condensates'', Phys. Rev. A {\bf 55} 4330 (1997).
\bibitem{fattori2008} M. Fattori, C. D'Errico,G. Roati, M. Zaccanti, M. Jona-Lasinio, M. Modugno, M. Inguscio and G. Modugno, ``Atom Interferometry with a Weakly Interacting Bose-Einstein Condensate'', Phys. Rev. Lett. {\bf 100} 080405 (2008).
\bibitem{Kasevich3} W. Li, A. K. Tuchman, H-C Chien and M. A. Kasevich, ``Extended Coherence Time with Atom-Number Squeezed States'', Phys. Rev. Lett. {\bf 98} 040402 (2007).
\bibitem{Jo2007} G-B Jo, Y. Shin, S. Will, T. A. Pasquini, M. Saba, W. Ketterle, D. E. Pritchard, M. Vengalattore and M. Prentiss, ``Long Phase Coherence Time and Number Squeezing of Two Bose-Einstein Condensates on an Atom Chip'', Phys. Rev. Lett. {\bf 98} 030407 (2007).
\bibitem{Yoshio2000} Y. Torii, Y. Suzuki, M. Kozuma, T. Sugiura, T. Kuga, L. Deng and E. W. Hagley, ``Mach-Zehnder Bragg interferometer for a Bose-Einstein condensate'', Phys. Rev. A {\bf 61} 041602(R) (2000).
\bibitem{Norman1950} N. F. Ramsey, ``A Molecular Beam Resonance Method with Separated Oscillating Fields'', Phys. Rev. {\bf 78} 695 (1950).
\bibitem{Monika2008} M. H. Schleier-Smith, I. D. Leroux and V. Vuleti\'c, Reduced-Quantum-Uncertainty States of an Ensemble of Two-Level Atoms, arXiv:0810.2582v1
\bibitem{J2008} J. Appel, P. J. Windpassinger, D. Oblak, U. B. Hoff, N. Kj{\ae}rgaard and E. S. Polzik, ``Mesoscopic atomic entanglement for precision measurements beyond the standard quantum limit'', arXiv:0810.3545v1
\bibitem{JE2008} J. Est$\grave{\mbox{e}}$ve, C. Gross, A. Weller, S. Giovanazzi and M. K. Oberthaler, ``Squeezing and entanglement in a BoseÐEinstein condensate'', Nature {\bf 455} 1216 (2008).
\bibitem{D2008} D. D\"oring, N. P. Robins, C. Figl and J. D. Close, ``Probing a Bose-Einstein condensate with an atom laser'', Opt. Express {\bf 16} 13893 (2008).
\bibitem{M1997} M-O. Mewes, M. R. Andrews, D. M. Kurn, D. S. Durfee, C. G. Townsend and W. Ketterle, ``Output Coupler for Bose-Einstein Condensed Atoms'', Phys. Rev. Lett. {\bf 78} 582 (1997).
\bibitem{JD2008} J. E. Debs, D. D\"oring, N. P. Robins, C. Figl, P. A. Altin and J. D. Close, ``A two-state Raman coupler for coherent atom optics'', Opt. Express {\bf 17} 2319 (2009).
\bibitem{P2003} P. J. Lee, B. B. Blinov, K. Brickman, L. Deslauriers, M. J. Madsen, R. Miller, D. L. Moehring, D. Stick and C. Monroe, ``Atomic qubit manipulations with an electro-optic modulator'', Opt. Lett. {\bf 28} 1582 (2003).
\bibitem{Immanuel2001} I. Bloch, M. K\"ohl, M. Greiner, T. W. H\"ansch and T. Esslinger, ``Optics with an Atom Laser Beam'', Phys. Rev. Lett. {\bf 87} 030401 (2001).
\bibitem{wynands} R. Wynands and S. Weyers,  ``Atomic fountain clocks'', Metrologia, {\bf 42} S64 (2005).
\bibitem{N2008} N. P. Robins, C. Figl, M. Jeppesen, G. R. Dennis and J. D. Close, ``A pumped atom laser'', Nat. Phys. {\bf 4} 731 (2008).
\bibitem{DD2008}D. D\"oring, G. R. Dennis, N. P. Robins, M. Jeppesen, C. Figl, J. J. Hope and J. D. Close, ``Pulsed pumping of a Bose-Einstein condensate'', Phys. Rev. A {\bf 79} 063630 (2009).


\end{thebibliography}
\end{document}